\documentclass[pra,aps,twocolumn,showpacs,superscriptaddress,floatfix,amsmath,amssymb,nofootinbib]{revtex4}
\newcommand{\beq}{\begin{equation}}
\newcommand{\eeq}{\end{equation}}
\newcommand{\beqa}{\begin{eqnarray}}
\newcommand{\eeqa}{\end{eqnarray}}
\def\ra{\rangle}
\def\la{\langle}
\usepackage{graphicx}
\usepackage{bm}

\usepackage{color}

\begin{document}
\title{Shortcuts to adiabaticity  by superadiabatic iterations}
\author{S. Ib\'a\~{n}ez}
\affiliation{Departamento de Qu\'{\i}mica F\'{\i}sica, Universidad del Pa\'{\i}s Vasco - Euskal Herriko Unibertsitatea,
Apdo. 644, Bilbao, Spain}
\author{Xi Chen} 
\affiliation{Departamento de Qu\'{\i}mica F\'{\i}sica, Universidad del Pa\'{\i}s Vasco - Euskal Herriko Unibertsitatea, 
Apdo. 644, Bilbao, Spain}
\affiliation{Department of Physics, Shanghai University, 200444 Shanghai, People's Republic of China}
\author{J. G. Muga}
\affiliation{Departamento de Qu\'{\i}mica F\'{\i}sica, Universidad del Pa\'{\i}s Vasco - Euskal Herriko Unibertsitatea,
Apdo. 644, Bilbao, Spain}
\affiliation{Department of Physics, Shanghai University, 200444 Shanghai, People's Republic of China}
\begin{abstract}
Different techniques to speed up quantum adiabatic processes
are currently being explored for applications in atomic, molecular and optical physics, such as transport, cooling and expansions, 
wavepacket splitting, or internal state control. Here we examine the 
capabilities of superadiabatic iterations to produce a sequence of 
shortcuts to adiabaticity. The general formalism is worked out as well as examples for population inversion in a two-level system.      
\end{abstract}
\maketitle
\section{Introduction}
There is currently much interest to speed up quantum adiabatic processes 
for applications such as fast cold-atom or ion transport,
expansions, wave-packet splitting or internal state 
population and state control \cite{review}. 
Different techniques have been put forward and/or applied.
Among them Demirplack and Rice \cite{DR03,DR05,DR08}, and Berry \cite{Berry2009} 
proposed the 
addition of a suitable counterdiabatic term $H_{cd}^{(0)}(t)$ 
to the time dependent Hamiltonian $H_0(t)$ whose adiabatic dynamics is to be implemented. With that term added 
transitions in the instantaneous eigenbasis $\{|n_0(t)\ra\}$
of $H_0(t)$ are suppressed,  $H_0(t) |n_0(t)\ra = E_n^{(0)}(t) |n_0(t)\ra$,
while there are in general transitions in the instantaneous eigenbasis of the new full Hamiltonian $H_0+H_{cd}^{(0)}$. Experiments that implement these ideas have been recently performed in different two-level systems \cite{Oliver,Zhang}. 
The same $H_{cd}^{(0)}(t)$ also appears naturally when studying the adiabatic approximation of the reference system, the one that evolves with $H_0(t)$, see e.g. \cite{Mes}.
The reference system behaves adiabatically, following the eigenstates of 
$H_0(t)$, when the counterdiabatic term is negligible, and the adiabatic approximation is close to the  actual dynamics.
This is made evident in an interaction picture  (IP)  based on the 
unitary transformation $A_0(t)=\sum_n |n_0(t)\ra\la n_0(0)|$.
(The ``parallel-transport'' condition $\la n_0(t)|\dot{n}_0(t)\ra=0$ is assumed hereafter 
to define the phases.)  
From the Schr\"odinger equation $i\hbar \partial_t \psi_0(t)=H_0(t)\psi_0(t)$
and defining $\psi_1(t)=A_0^\dagger\psi_0(t)$,
the IP  
equation $i\hbar \partial_t \psi_1(t)=H_1(t)\psi_1(t)$
is deduced, where 
$H_1(t)=A_0^\dagger(t)(H_0(t)-K_0(t))A_0(t)$ is the effective 
IP Hamiltonian and $K_0(t)=i\hbar \dot{A}_0(t)A_0^\dagger(t)$
is a coupling term. 
If $K_0(t)$ is zero or negligible, $H_1(t)$ becomes diagonal in the basis 
$\{|n_0(0)\ra\}$, so that the IP equation becomes an uncoupled system
with solutions 
\beq
|\psi_1(t)\ra=U_1(t) |\psi_1(0)\ra,
\eeq
where
\beq
U_1(t)=\sum_n |n_0(0)\ra e^{-\frac{i}{\hbar}\int_0^t E_n^{(0)}(t')dt'}
\la n_0(0)| 
\eeq
is the unitary evolution operator for the uncoupled system.
Correspondingly, from $|\psi_0(t)\ra= A_0(t) |\psi_1(t)\ra$,
\beq
|\psi_0^{(1)}(t)\ra=\sum_n |n_0(t)\ra e^{-\frac{i}{\hbar}\int_0^t E_n^{(0)}(t')dt'}
\la n_0(0)|\psi_0(0)\ra, 
\eeq
where we have used $|\psi_1(0)\ra=|\psi_0(0)\ra$ since $A_1(0)=1$ by construction.   
The same solution, which, for a non-zero $K_0(t)$, is only approximate, may become  
exact by adding to the IP Hamiltonian 
the counterdiabatic term $A_0^\dagger(t)K_0(t)A_0(t)$. 
This requires an external intervention and changes the physics of the  original system,
so that $U_1(t)$ describes the evolution exactly.  
In the IP the modified Hamiltonian is $H^{(1)}(t)=H_1(t)+A_0^\dagger(t)K_0(t)A_0(t)=
A_0^\dagger(t)H_0(t)A_0(t)$, and in  
the Schr\"odinger picture (SP)
the additional term becomes simply $K_0$. The modified Schr\"odinger Hamiltonian is $H_0^{(1)}(t)=H_0(t)+K_0(t)$, so  we identify $H_{cd}^{(0)}(t)=K_0(t)$.
A ``small'' coupling term $K_0(t)$ that makes the adiabatic approximation a good one also implies a small  
counterdiabatic manipulation but, irrespective of the size of $K_0(t)$,  $H_0^{(1)}(t)$ provides a shortcut to slow adiabatic following because 
it keeps the populations in the instantaneous basis of $H_0(t)$ invariant, 
in particular at the final time $t_f$. Moreover, if $K_0(0) = 0$ and $K_0(t_f) = 0$, then $H_0^{(1)}=H_0$ at $t=0$ and $t=t_f$.
This is useful in practice to ensure the continuity of the Hamiltonian 
at the boundary times: usually  $H_0(t<0)=H_0(0)$ and $H_0(t>t_f)=H_0(t_f)$, 
so $K_0(t<0)=K_0(t>t_f)=0$, i.e., $H_0(t)$ is the actual Hamiltonian before and after the process. 

The previous formal framework   
may be repeated iteratively to define further IPs 
by diagonalizing the 
effective Hamiltonians of each IP.     
These iterations were used to 
establish {\it generalized adiabatic invariants} and {\it adiabatic invariants 
of $n$-th order} by Garrido \cite{Garrido}. Berry also used 
this iterative procedure to calculate a sequence of corrections to Berry's phase for cyclic processes with finite slowness,
and introduced the concept of ``superadiabaticity'' \cite{Berry1987}.  
For later developments and applications see e.g. \cite{Berry1990,pertur,Joli,DR08,master,MagReson,Berry&Uzdin,Uzdin&Moiseyev,Moiseyev,Oliver,Sara12}. 

The idea of superadiabatic iterations is best understood by working 
out explicitly the next interaction picture:\footnote{The first IP and iteration just described, 
with dynamics governed by  
$H_1(t)$, generates the modified dynamics based on $H_0^{(1)}$ in the SP. This iteration may be naturally   
termed as ``adiabatic'', since the unitary transformation used, $A_0$, relies on the usual adiabatic basis. Moreover this is the IP used to perform the adiabatic approximation by 
neglecting $K_0$. The second iteration may be considered as the first ``superadiabatic''
one.}
let us start with 
$i\hbar\partial_t \psi_1(t)=H_1(t)\psi_1(t)$ and treat it as if it were, formally, 
a Schr\"odinger equation. The diagonalization of $H_1(t)$ 
provides the eigenbasis $\{|n_1(t)\ra\}$, {$H_1(t)|n_1(t)\ra = E_n^{(1)}(t)|n_1(t)\ra$},
that we fix again with the parallel transport condition, $\la n_1(t)|\dot{n}_1(t)\ra=0$.
A new unitary operator $A_1=\sum_n|n_1(t)\ra\la n_1(0)|$ plays now the same 
role as $A_0$ in the first (adiabatic) IP. It defines a new interaction picture wave function $\psi_2(t)=A_1^\dagger(t)\psi_1(t)$ that satisfies 
$i\hbar\partial_t \psi_2(t)=H_2(t)\psi_2(t)$, where $H_2(t)=A_1^\dagger(t)(H_1(t)-K_1(t))A_1(t)$ and $K_1=i\hbar\dot{A}_1A_1^\dagger$. 
If $K_1$ is zero or ``small'' enough, i.e., if a (first order) 
superadiabatic approximation is valid, 
the dynamics would be uncoupled in the new interaction picture, namely, 
\beq
|\psi_2(t)\ra=U_2(t)|\psi_2(0)\ra,
\eeq
where
\beq
U_2(t)= \sum_n |n_1(0)\ra e^{-\frac{i}{\hbar}\int_0^t E_n^{(1)}(t')dt'}
\la n_1(0)|
\eeq
is the approximate evolution operator in the second IP for uncoupled motion.
It may happen that a process is not adiabatic, 
since $K_0(t)$ may not be neglected, but (first-order) superadiabatic when $K_1(t)$ can be neglected. Transforming back to 
the Schr\"odinger picture, $|\psi_0^{(2)}(t)\ra = A_0(t) A_1(t) U_2(t) |\psi_2(0)\ra$ becomes 
\beqa
|\psi_0^{(2)}(t)\ra &=& \sum_n\sum_m |m_0(t)\ra \la m_0(0)|n_1(t)\ra e^{-\frac{i}{\hbar}\int_0^t E_n^{(1)}(t')dt'} 
\nonumber \\
&\times&  \la n_1(0)|\psi_0(0)\ra, 
\label{psi02}
\eeqa
%
%
and   
$|\psi_2(0)\ra=|\psi_0(0)\ra$ since $A_0(0)=A_1(0)=1$. 
Garrido distinguished two different aspects \cite{Garrido}: 
\begin{itemize}
\item Generalized adiabaticity: The evolution operator $A_0(t) A_1(t) U_2(t)$ provides an approximation to the actual (Schr\"odinger) dynamics
up to a correction term of order $1/t_f^{2}$. This is so without imposing any boundary conditions (BCs) at $t=0$ and $t=t_f$ on the Hamiltonian $H_0$.   
\item Higher order adiabaticity: $A_0(t) A_1(t) U_2(t)$ does not guarantee in general that 
$|n_0(0)\ra$ evolves into $|n_0(t_f)\ra$, up to a phase factor. If this is the  objective, in other words, if a superadiabatic approximation should behave, at final times, like the adiabatic approximation, up to phase factors, 
then some BCs have to be imposed. Garrido discussed how generalized adiabaticity implies higher order adiabaticity when BCs at the 
boundary times   are imposed on the derivatives of $H_0$.  
\end{itemize}
(Garrido's distinction does not apply in \cite{Berry1987} since there, 
it is assummed from the start that all the derivatives of the Hamiltonian $H_0$ vanish at the 
(infinite) time edges.)  
The second aspect  is crucial to design shortcuts to adiabaticity
for finite process times using the superadiabatic iterative structure, so let us 
be more specific.  
First notice that Eq. (\ref{psi02}) becomes exact if the term $A_1^{\dag} K_1 A_1$ is added to the IP Hamiltonian, so that now the modified IP Hamiltonian is
$H^{(2)} = H_2+A_1^{\dag} K_1 A_1= A_1^{\dag} H_1 A_1$. Then the modified SP Hamiltonian becomes $H_0^{(2)}(t)=H_0(t)+H_{cd}^{(1)}(t)$, where
$H_{cd}^{(1)}(t) = A_0(t) K_1(t) A_0^{\dag}(t)$.
However,   
quite generally the populations of the final state (\ref{psi02}) 
in the adiabatic basis $\{|n_0(t_f)\ra\}$ will be different 
from the ones of the adiabatic process, unless
(a) $\{|n_0(0)\ra\}  = \{|n_1(t_f)\ra\}$, up to phase factors, 
and (b) $\{|n_0(0)\ra\} =\{|n_1(0)\ra\}$, also up to phase factors. 
(a) is satisfied 
when $K_0(t_f)=0$. This makes $H_1(t_f)$ diagonal in the basis $\{|n_0(0)\ra \}$
and $[H_1(t_f),H_0(0)]=0$. 
(b) is satisfied when  
$K_0(0)=0$, which implies $H_1(0)=H_0(0)$.
In summary the requirement is that the eigenstates of $H_1(t)$ at $t=0$ and $t=t_f$ coincide with the eigenstates of $H_0(0)$. 
If, in addition, $K_1(0)=K_1(t_f)=0$,  
not only the populations, but also the initial and final Hamiltonians are the same
for the ``corrected'' and for the reference processes,
namely $H_0^{(2)}(0)= H_0(0)$ and $H_{0}^{(2)} (t_f)=  H_0(t_f)$.
   
Further iterations define higher order
superadiabatic frames with IP equations $i\hbar\partial_t\psi_j(t)=H_j\psi_j(t)$, where
\beqa
H_j&=&A_{j-1}^\dagger (H_{j-1}-K_{j-1})A_{j-1},
\label{Hj}
\\
K_j&=&i\hbar\dot{A}_jA_j^\dagger,
\label{Kj}
\eeqa
with $A_j(t)= \sum_n |n_j(t)\ra \la n_j(0)|$ and $H_j(t) |n_j(t)\ra = E_n^{(j)}(t) |n_j(t)\ra$. 
As the $A_j(0)=1$ by construction, there is a common initial state  $| \psi_j(0) \ra=| \psi_0(0) \ra$ for all iterations.
The general form for the modified IP Hamiltonians is $H^{(j)} = H_j+A_{j-1}^{\dag} K_{j-1} A_{j-1}= A_{j-1}^{\dag} H_{j-1} A_{j-1}$. Thus, the form of the modified Hamiltonians in the SP is 
\beq
H_0^{(j)} = H_0+H_{cd}^{(j-1)},
\eeq
where the SP counterdiabatic term is 
\beq
\label{Hcd}
H_{cd}^{(j)} = B_j K_j B_j^\dagger =  i\hbar B_j \dot{A}_j A_j^\dagger B_j^\dagger,
\eeq
with $B_j = A_0 \cdot\cdot\cdot A_{j-1}$ and $B_0=I$.
If $K_{j-1}(t)$ is small or negligible, $H_j(t)$ becomes diagonal in the basis $\{ |n_{j-1}(0)\ra \}$ and the IP equation becomes an uncoupled system with solutions $|\psi_j(t)\ra = U_j(t) |\psi_j(0)\ra$, where
\beq
U_j(t)= \sum_n |n_{j-1}(0)\ra e^{-\frac{i}{\hbar}\int_0^t E_n^{(j-1)}(t')dt'}
\la n_{j-1}(0)|
\eeq
is the approximate evolution operator in the $j$-th IP.
Correspondingly, the approximate solution in the SP is given by
$|\psi_0^{(j)}(t)\ra = A_0(t) A_1(t) \cdot \cdot \cdot A_{j-1}(t) U_j(t) |\psi_0(0)\ra$. This solution becomes exact if the $H_{cd}^{(j-1)}$ term is added to $H_0$, where in general the populations of $|\psi_0^{(j)}(t_f)\ra$ in the adiabatic basis $\{|n_0(t_f)\ra\}$ will be different from the ones of the adiabatic process, unless appropriate BCs are impossed.
These boundary conditions are made explicit in the next section, and correspond partially to the 
conditions discussed by Garrido in \cite{Garrido} 
to define ``higher order adiabaticity''.

Is there any advantage in using one or another counter-diabatic scheme?
There are several reasons that could make higher order schemes attractive
in practice: one is that the structure of the $H_{cd}^{(j)}$ may change with $j$. For example, for a two-level atom population inversion problem,
$H_{cd}^{(0)}=\hbar(\dot{\Theta}_0/2)\sigma_y$, whereas 
$H_{cd}^{(1)}=\hbar(\dot{\Theta}_1/2)(\cos\Theta_0 \sigma_x-\sin\Theta_0\sigma_z)$,
where the $\Theta_j$ are the polar angles corresponding to the Cartesian components of the Hamiltonian $H_j$, and the $\sigma_u$, with $u=x,y,z$, are  Pauli matrices  \cite{Sara12}. (We shall use the Cartesian decomposition $X\sigma_x+Y\sigma_y+Z\sigma_z$
for different Hamiltonians below.)

A second reason is that, for a fixed process time, the cd-terms tend to be  
smaller in norm as $j$ increases, up to a value
in which they begin to grow \cite{MagReson}. An optimal iteration may thus be set 
\cite{Berry1990,MagReson}. The ``asymptotic character'' of the superadiabatic 
coupling terms and the eventual divergence of the sequence can be traced back to
the existence of non-adiabatic transitions, even if they are small \cite{Berry1987}.

To generate shortcuts one should pay attention though not only to the size of the cd-terms but also to the feasibility or approximate fulfillment of the 
required BCs at the boundary times. 
Thus, it may happen that an ``optimal iteration'', of minimal 
norm for the cd-term, fails to provide a shortcut because of the
BCs, as illustrated below in Sect. IV.      
\section{Boundary conditions for shortcuts to adiabaticity via superadiabatic 
iterations}
In this section we set the boundary conditions that guarantee that 
$H_0^{(j)}(t)$ provides a shortcut to adiabaticity. 
We have seen that for $j=1$ no conditions are required. 
For $j=2$ we need that $\{|n_{1}(t_f)\ra \}=\{|n_0(0)\ra \}$ and $\{|n_{1}(0)\ra \}=\{|n_0(0)\ra \}$, (as before in these and similar expressions in brackets, 
the equalities should be understood up to phase factors), i.e.,  
$K_0(t_f)=K_0(0)=0$.
For the iterations $j>2$ we need that
(a) $\{|n_{j-1}(0)\ra \}=\{|n_0(0)\ra \}$, 
which occurs when  $K_{j-2}(0)=K_{j-3}(0)= ... =K_1(0)=K_0(0)=0$, and
(b) $\{|n_{j-1}(t_f)\ra \}=\{|n_{j-2}(0)\ra \}$, $\{|n_{j-2}(t_f)\ra \}=\{|n_{j-3}(0)\ra \}$, $\{|n_{j-3}(t_f)\ra \}=\{|n_{j-4}(0)\ra \}$, ... , and $\{|n_1(t_f)\ra \}=\{|n_0(0)\ra \}$.
This amounts to imposing  
$K_{j-2}(t_f)=K_{j-3}(t_f)= ... =K_1(t_f)=K_0(t_f)=0$. 
The vanishing of $K_{j'}(0)$ for $j'\leq j-2$ implies that $H_0(0)=H_1(0)=...=H_{j-1}(0)$, so  
(a) and (b) combined may be summarized as 
$\{|n_{j'}(0)\ra \}=\{|n_{j'}(t_f)\ra \}=\{|n_0(0)\ra \}$
for all $j'\leq j-1$.  
Garrido 
showed that canceling out the first $l$-th time derivatives of $H_0(0)$ and $H_0(t_f)$ makes $K_j(0)=0$ and $K_j(t_f)=0$, for $j=1, ... ,l-1$, respectively \cite{Garrido}. However canceling out the derivatives of $H_0$ is a sufficient but not a necessary 
condition to cancell the coupling terms, so 
we find it more useful to focus instead  on the coincidence of the bases, this is  
exemplified in Sect. IV.  
%
%
%
%
%
%
%
%
%
%
\section{Alternative framework with a constant basis}
An alternative to the formal framework described so far provides  computational
advantages.  
It was implicity applied by Demirplak and Rice for a two-level system
\cite{DR08}.
We shall here generalize and formulate explicitly this approach and show its essential equivalence to the former.
The main idea is to use instead of the $A_j$ a different set of 
unitary operators, $\tilde{A}_j(t) = \sum_{n} |\tilde{n}_j(t)\ra \la n|$, to define the sequence of interaction pictures, where  $|\tilde{n}_j(t)\ra$ 
are eigenstates of the new IP Hamiltonians  $\tilde{H}_{j}(t)$, {such that $\tilde{H}_{j}(t)  |\tilde{n}_j(t)\ra =  \tilde{E}_n^{(j)}(t) |\tilde{n}_j(t)\ra$,}
and $\{ |n\ra \}$ is a constant orthonormal basis {\it equal for all} $j$, which in principle does not necessarily coincide with $|n_j(0)\ra$. 
Similarly to Eq. (\ref{Hj}), 
%
\beq
\label{H_j+1'}
\tilde{H}_{j} = \tilde{A}_{j-1}^\dagger(\tilde{H}_{j-1}-\tilde{K}_{j-1})\tilde{A}_{j-1},
\eeq
where 
$\tilde{K}_j = i\hbar \dot{\tilde{A}}_j \tilde{A}_j^\dagger = i \hbar \sum_{n} |\dot{\tilde{n}}_j(t)\ra\la \tilde{n}_j(t)|$.
The counterdiabatic terms in the SP are introduced as before, 
$\tilde{H}_{cd}^{(j)} = \tilde{B}_j \tilde{K}_j \tilde{B}_j^\dagger$, where $\tilde{B}_j = \tilde{A}_0 \cdot\cdot\cdot \tilde{A}_{j-1}$ with
$\tilde{B}_0=I$.
We shall next show that these cd-terms are independent of the chosen constant basis, so that   
$\tilde{H}_{cd}^{(j)}(t) = H_{cd}^{(j)}(t)$. 
Therefore, it is worth using $\tilde{A}_j(t)$ instead of $A_j(t)$ since they are  simpler operators and significantly facilitate the manipulations  as a common basis is used. 

Let us start with the first iteration. Since $\tilde{H}_{0}(t)=H_{0}(t)$, then $\tilde{E}_n^{(0)}(t)= E_n^{(0)}(t)$,
$|\tilde{n}_0(t)\ra = |n_0(t)\ra$, and
$\tilde{K}_0=K_0$, so $\tilde{H}_{cd}^{(0)}=H_{cd}^{(0)}$. In addition, from Eq. (\ref{Hj}),
$H_0-K_0 = A_0 H_{1}A_0^\dagger$, and substituting it in Eq. (\ref{H_j+1'}) leads to
\beq
\label{H_1'}
\tilde{H}_1= u_0 H_1 u_0^\dagger,
\eeq
where we have defined a constant unitary operator
\beqa
u_0 &=& \tilde{A}_0^\dagger A_0 = \sum_{n} |n\ra \la n_0(0)|,
\nonumber
\\
\dot{u}_0 &=& 0.
\nonumber
\eeqa
Using  
\beq
\label{H_E}
H_{j}(t) = \sum_{n} |n_{j}(t)\ra E_n^{(j)}(t) \la n_{j}(t)|
\eeq
and
\beq
\label{H_E'}
\tilde{H}_{j}(t) = \sum_{n} |\tilde{n}_{j}(t)\ra \tilde{E}_n^{(j)}(t) \la \tilde{n}_{j}(t)|,
\eeq
for $j=1$ in Eq. (\ref{H_1'}), we get that $\tilde{E}_n^{(1)}(t) = E_n^{(1)}(t)$ and $|\tilde{n}_{1}(t)\ra = u_0 |n_{1}(t)\ra$, while $|n\ra = u_0 |n_{0}(0)\ra$. Expanding $\tilde{H}_{cd}^{(1)} = \tilde{A}_0 \tilde{K}_1 \tilde{A}_0^\dagger$ we have that
\beqa
\tilde{H}_{cd}^{(1)}(t) = i \hbar \sum_{n,m,l,p} |\tilde{n}_0(t)\ra  \la n|\dot{\tilde{m}}_1(t)\ra \la m|l\ra \la \tilde{l}_1(t)|p\ra \la \tilde{p}_0(t)|. 
\nonumber
\eeqa
Using now $\la m|l\ra = \delta_{m l}$, $|\tilde{n}_0(t)\ra = |n_0(t)\ra$, $|n\ra = u_0 |n_{0}(0)\ra$, and
$|\tilde{n}_{1}(t)\ra = u_0 |n_{1}(t)\ra$, it follows that $\tilde{H}_{cd}^{(1)}=H_{cd}^{(1)}$. Also, 
$\tilde{K}_1 = \tilde{A}_0^\dagger A_0 K_1 A_0^\dagger \tilde{A}_0 = u_0 K_1 u_0^\dagger$.

Repeating these steps for  $j\geqslant1$, $\tilde{H}_{j} = u_{j-1} H_{j} u_{j-1}^\dagger$ and
$\tilde{K}_{j} = u_{j-1} K_{j} u_{j-1}^\dagger$, where 
%
\beqa
u_j&=& \tilde{A}_j^\dagger u_{j-1} A_j = \sum_{n} |n\ra \la n_j(0)|,
\nonumber
\\
\dot{u}_j&=& 0.
\nonumber
\eeqa
This leads to $\tilde{E}_n^{(j)}(t) = E_n^{(j)}(t)$, $|\tilde{n}_{j}(t)\ra = u_{j-1} |n_{j}(t)\ra$, and $|n\ra = u_{j-1} |n_{j-1}(0)\ra$.
Thus, for all $j\geq 0$,
\beq
\tilde{H}_{cd}^{(j)} = H_{cd}^{(j)}.
\nonumber
\eeq
The boundary conditions to achieve shortcuts to adiabaticity
take the same form as for the original framework in the previous section.  
Since  $\tilde{K}_{0} = K_{0}$ and
$\tilde{K}_{j} = u_{j-1} K_{j} u_{j-1}^\dagger$ for $j \geqslant 1$, 
for the $j$-th iteration, {with $j>1$}, we need that $\tilde{K}_0(0)=\tilde{K}_1(0)=...=\tilde{K}_{j-2}(0)=0$, and
$\tilde{K}_0(t_f)=\tilde{K}_1(t_f)=...=\tilde{K}_{j-2}(t_f)=0$. 
Let us recall that no conditions were required for $j=1$, although, as shown 
in the next section, using a convenient (constant or initial adiabatic) basis 
for specific Hamiltonians may also lead to conditions for $j=1$.   
\section{Two-level atom}
The general formalism will now be applied to the two-level atom. Assuming a semiclassical interaction between a laser electric field and the atom, the electric dipole and the rotating wave approximations, the Hamiltonian of the system in a laser-adapted IP (that plays the role of the Schr\"{o}dinger picture of the previous section)  is
\beq   
H_{0}(t)=\frac{\hbar}{2}
\left(\begin{array}{cc} 
-\Delta(t) & \Omega_{R}(t)
\\
\Omega_{R}(t) & \Delta(t)
\end{array} \right),
\label{H0_2level}
\eeq
where $\Omega_{R}(t)$ is the Rabi frequency, assumed real, and $\Delta(t)$ is the detuning, in the ``bare basis'' of the two level system, $|1\rangle = (\tiny{\begin{array} {c} 1\\ 0 \end{array}})$, $|2\rangle =
(\tiny{\begin{array} {c} 0\\ 1 \end{array}})$. The Hamiltonians of the consecutive interaction pictures can be written as \cite{DR08}
\beq
\tilde{H}_j(t)=\left(
\begin{array}{cc}
Z_j(t) & X_j(t)-iY_j(t)
\\
X_j(t)+iY_j(t)& -Z_j(t)
\end{array}
\right),
\label{Hj_2level}
\eeq
or $\tilde{H}_j=X_j\sigma_x+Y_j\sigma_y+Z_j\sigma_z$  \cite{Sara12}. Then, $X_0(t)=\hbar \Omega_{R}(t)/2$, $Y_0(t)=0$ and $Z_0(t)=- \hbar \Delta(t)/2$. $X_j$, $Y_j$, and $Z_j$ are the Cartesian coordinates of the ``trajectory'' of $\tilde{H}_j(t)$. It is also useful to consider the corresponding polar, azimuthal and radial spherical coordinates, $\Theta_j(t)$, $\Phi_j(t)$, and $R_j(t)$  \cite{DR08,Sara12}, that satisfy 
\beqa
\label{spherical}
\cos(\Theta_j) &=& \frac{Z_j}{R_j} ,\,\,\,\,  \sin(\Theta_j)= \frac{P_j}{R_j} ,\,\,\,\,  0 \leq \Theta_j \leq \pi,
\nonumber
\\
\cos(\Phi_j) &=&  \frac{X_j}{P_j} ,\,\,\,\,  \sin(\Phi_j)= \frac{Y_j}{P_j} ,\,\,\,\,  0 \leq \Phi_j \leq 2\pi,   \,\,\,\,\,\,\,\,
\eeqa
with $R_j= \sqrt{X_j^2+Y_j^2+Z_j^2}$ and $P_j= \sqrt{X_j^2+Y_j^2}$, where the positive branch is taken.
The eigenvalues of $\tilde{H}_j(t)$ are $E_1^{(j)}=-R_j$ and $E_2^{(j)}=R_j$, and
the corresponding eigenstates $\{|\tilde{n}_j(t)\ra \}$ are 
%
\beqa
\label{eigenstates}
|\tilde{1}_{j}\ra &=& e^{i \varepsilon_j} \left[  e^{-i \Phi_j /2} \sin{\left(\frac{\Theta_j}{2}\right)} |1\ra
- e^{i \Phi_j /2} \cos{\left(\frac{\Theta_j}{2}\right)} |2\ra \right],
\nonumber
\\
|\tilde{2}_{j}\ra &=& e^{-i \varepsilon_j} \left[  e^{-i \Phi_j /2} \cos{\left(\frac{\Theta_j}{2}\right)} |1\ra
+ e^{i \Phi_j/2} \sin{\left(\frac{\Theta_j}{2}\right)} |2\ra \right],
\nonumber
\\
\eeqa
where the phase 
\beq
\label{epsilon}
\varepsilon_j(t)= - \frac{1}{2} \int_0^t \dot{\Phi}_j(t') \cos{[\Theta_j(t')]} dt'
\eeq
is introduced to fulfill the parallel transport condition $\la \tilde{n}_j|\dot{\tilde{n}}_j \ra=0$.
We define $\tilde{A}_j=|\tilde{1}_j(t)\ra\la 1|+|\tilde{2}_j(t)\ra\la 2|$. 
The matrix $\tilde{A}_j(t)$ under these conditions is
\beq
\tilde{A}_j=\left(
\begin{array}{cc}
\sin{\left(\frac{\Theta_j}{2}\right)} e^{i\varepsilon_j-i\Phi_j /2} & \cos{\left(\frac{\Theta_j}{2}\right)}
e^{-i\varepsilon_j-i\Phi_j /2}
\\
-\cos{\left(\frac{\Theta_j}{2}\right)} e^{i\varepsilon_j+i\Phi_j /2} & \sin{\left(\frac{\Theta_j}{2}\right)}
e^{-i\varepsilon_j+i\Phi_j/2} 
\end{array}
\right).
\label{Aj}
\eeq
Then, from Eq. (\ref{Kj}),
\beqa
\label{K_j}
\tilde{K}_j &=& \frac{\hbar}{2} \left[ -\dot{\Theta}_j \sin{\left(\Phi_j \right)}- \frac{\dot{\Phi}_j}{2} \cos{\left( \Phi_j \right)} \sin{\left(2\Theta_j \right)} \right] \sigma_x
\nonumber
\\
&+& \frac{\hbar}{2} \left[ \dot{\Theta}_j \cos{\left(\Phi_j \right)}- \frac{\dot{\Phi}_j}{2} \sin{\left( \Phi_j \right)} \sin{\left(2\Theta_j \right)} \right] \sigma_y
\nonumber
\\
&+& \frac{\hbar \dot{\Phi}_j}{2} \sin^2{\left(\Theta_j \right)} \sigma_z.
\eeqa
Note that $\tilde{A}_j^\dagger \tilde{K}_j \tilde{A}_j = \tilde{A}_j^\dagger \dot{\tilde{A}}_j$ has only non-diagonal elements in the bare basis
$\{|1\ra, |2\ra \}$ \cite{DR08}.

From Eq. (\ref{Hj}), the Cartesian coordinates of $\tilde{H}_{j+1}(t)$ are
\beqa
\label{XYZ}
X_{j+1} &=& \frac{\hbar}{2} \left[ \dot{\Theta}_{j} \sin{\left(2\varepsilon_{j} \right)}- \dot{\Phi}_{j} \sin{\left( \Theta_{j} \right)}
\cos{\left(2\varepsilon_{j} \right)} \right],
\nonumber
\\
Y_{j+1} &=& \frac{\hbar}{2} \left[ -\dot{\Theta}_{j} \cos{\left(2\varepsilon_{j} \right)}- \dot{\Phi}_{j} \sin{\left( \Theta_{j} \right)}
\sin{\left(2\varepsilon_{j} \right)} \right],
\nonumber
\\
Z_{j+1} &=& -R_{j}.
\eeqa
In general, if $\Phi_j(t)$ is constant for a particular $j=J$, then $\dot{\Phi}_J(t)=0$, and from Eq. (\ref{epsilon}), $\varepsilon_J(t)=0$. Thus, taking into account Eq. (\ref{XYZ}), we have that $X_{J+1}(t)=0$ and $Y_{J+1}(t)=-\hbar \dot{\Theta}_J/2$. Equation (\ref{spherical}) leads to $\Phi_{J+1}(t)= \{ \pi/2,3\pi/2 \}$, with $\pi/2$ when $Y_{J+1} > 0$
($\dot{\Theta}_J < 0$), and $3\pi/2$ when $Y_{J+1} < 0$ ($\dot{\Theta}_J > 0$).
If $Y_{J+1}=0$, $\Phi_{J+1}$ is discontinuous, and  $\Theta_{J+1}=\pi$.
Therefore,  $\varepsilon_{J+1}(t)=  \{ 0,\pm \pi/2 \}$.
From here, several general conditions can be deduced for $j' >J$:
$\Phi_{j'>J}(t)= \{ \pi/2,3\pi/2 \}$, $\varepsilon_{j' > J}(t)=  \{ 0,\pm \pi/2 \}$, $X_{j'>J}(t)=0$, and
$Y_{J+1}(t)=-\hbar \dot{\Theta}_J/2$ or $Y_{j' > J+1}(t)= \pm \hbar \dot{\Theta}_{j'-1}/2$.
Moreover, from Eq. (\ref{K_j}), 
$\tilde{K}_{j'>J}= \pm (\hbar \dot{\Theta}_{j'} /2) \sigma_x$ with positive sign if $\Phi_{j'}(t)= 3\pi/2$ and negative sign if $\Phi_{j'}(t)= \pi/2$.
Eq. (\ref{spherical}) and $Y_0(t)=0$ imply $\Phi_0(t)=0$ if $X_0(t)>0$ and $\Phi_0(t)=\pi$ if $X_0(t)<0$. We may thus take $J=0$ and apply the above relations, for example  $\dot{\Phi}_0(t)=0$ and  $\varepsilon_0(t)=0$.\footnote{The analysis in this paragraph follows closely  
\cite{DR08}, but some of the results differ, in particular the values allowed  
for the phases $\varepsilon_{j'>J}$.}

As we mentioned before, the method fails as a shortcut to adiabaticity when the boundary conditions are not well fulfilled. In order to have a shortcut generated by the iteration {$j$} we require that 
$\Delta(t)$ and $\Omega_R(t)$ are such that
%
{
\beqa
\label{IC}
|\tilde{1}_{j'}(0)\ra &\approx&  |1\ra  ,\,\,\,\,\,\,\,\,  |\tilde{2}_{j'}(0)\ra \approx |2\ra,
\\
|\tilde{1}_{j'}(t_f)\ra &\approx&  |1\ra  ,\,\,\,\,\,\,\,\,  |\tilde{2}_{j'}(t_f)\ra \approx |2\ra, 
\label{IC2}
\eeqa
for $0<j'< j$, up to phase factors.}
For $j'=0$ a natural and simple assumption is that the bare basis
coincides initially with the adiabatic
basis, i.e., Eq. (\ref{IC}); at $t_f$ we assume that the bare and adiabatic bases
also coincide, allowing for 
permutations in the indices and phase factors. 
  
At $t=0$, using Eq. (\ref{eigenstates}), taking into account that,
from Eq. (\ref{epsilon}), $\varepsilon_{j'}(0)=0$, and that $\Phi_0(t)=0$, 
$\sin{\left[\Theta_{j'}(0)/2\right]}=1$ and $\cos{\left[\Theta_{j'}(0)/2\right]}=0$
are required, or $\Theta_{j'}(0)=\pi$.
Then, $\cos{[\Theta_{j'}(0)]}= Z_{j'}(0)/R_{j'}(0)= -1$. This condition is fulfilled if
\beq
\label{cond_1}
Z_{j'}^2(0) \gg X_{j'}^2(0)+Y_{j'}^2(0),
\eeq
as long as  $Z_{j'=0}(0)<0$, and knowing that $Z_{j'>0}(t)=-R_{j'-1}(t)<0$. The condition (\ref{cond_1}) can be simplified for specific $j'$-values as
\beqa
|Z_0(0)| &\gg& |X_0(0)|,
\label{con00}
\\
|Z_{j'>0}(0)| &\gg& |Y_{j'>0}(0)|.
\label{cond_1_0}
\eeqa
At $t=t_f$, 
\beq
Z_{j'}^2(t_f) \gg X_{j'}^2(t_f)+Y_{j'}^2(t_f)
\eeq
should be satisfied, 
where now, $\Theta_0(t_f)$ can be either $0$, if $Z_0(t_f)>0$, or $\pi$ if $Z_0(t_f)<0$, and $\Theta_{j'>0}(t_f)=\pi$.
As before, this condition splits into
\beqa
|Z_0(t_f)| &\gg& |X_0(t_f)|,
\label{con0tf}
\\
|Z_{j'>0}(t_f)| &\gg& |Y_{j'>0}(t_f)|.
\label{cond_1_tf}
\eeqa
%

\begin{figure}[h]
\begin{center}
\includegraphics[height=4.cm,angle=0]{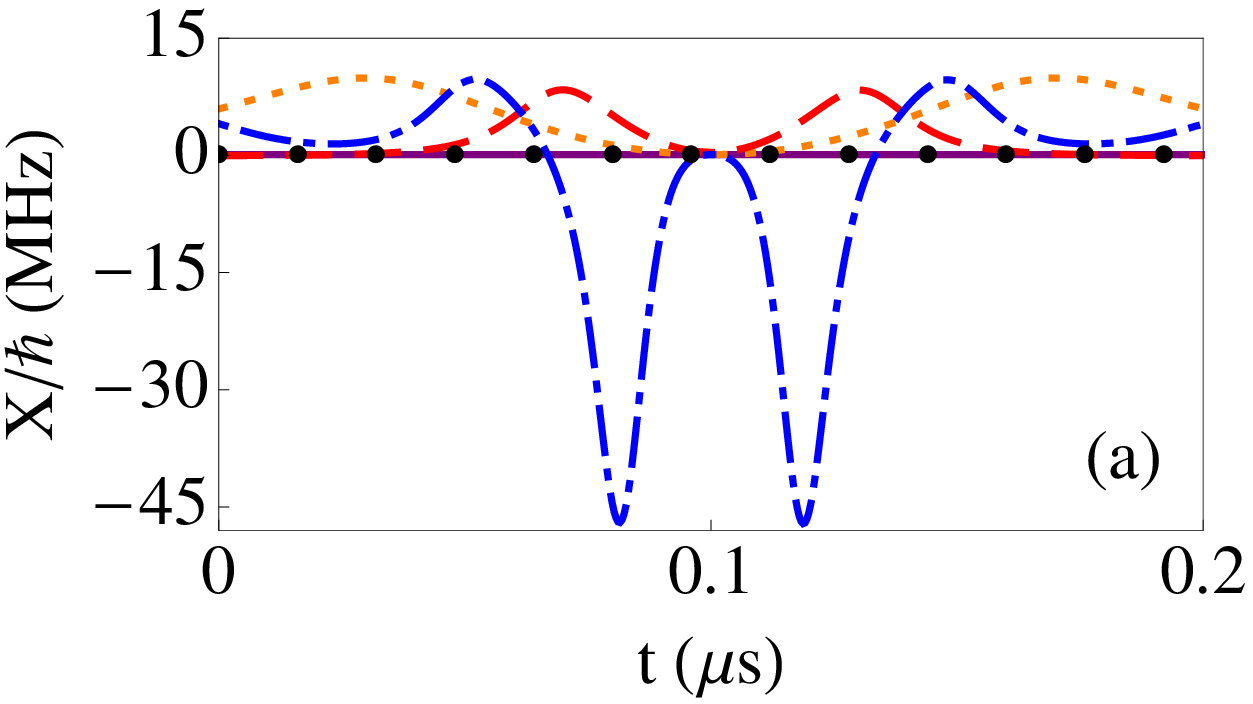}
\includegraphics[height=4.cm,angle=0]{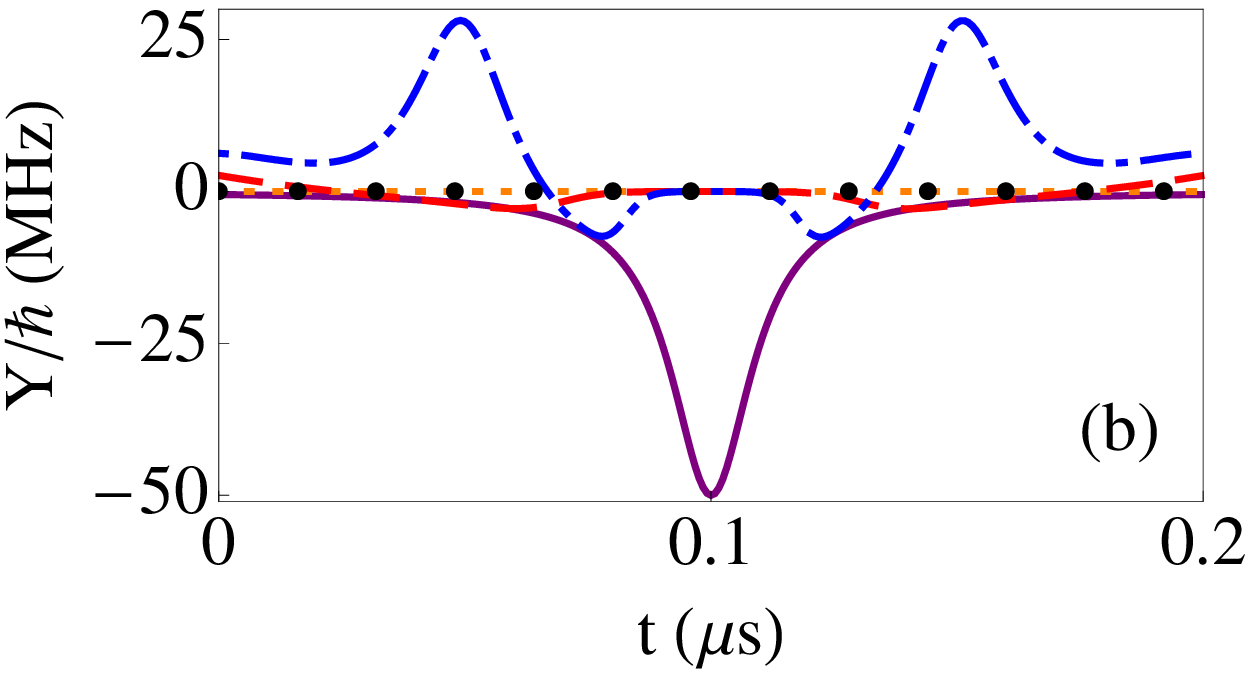}
\includegraphics[height=4.cm,angle=0]{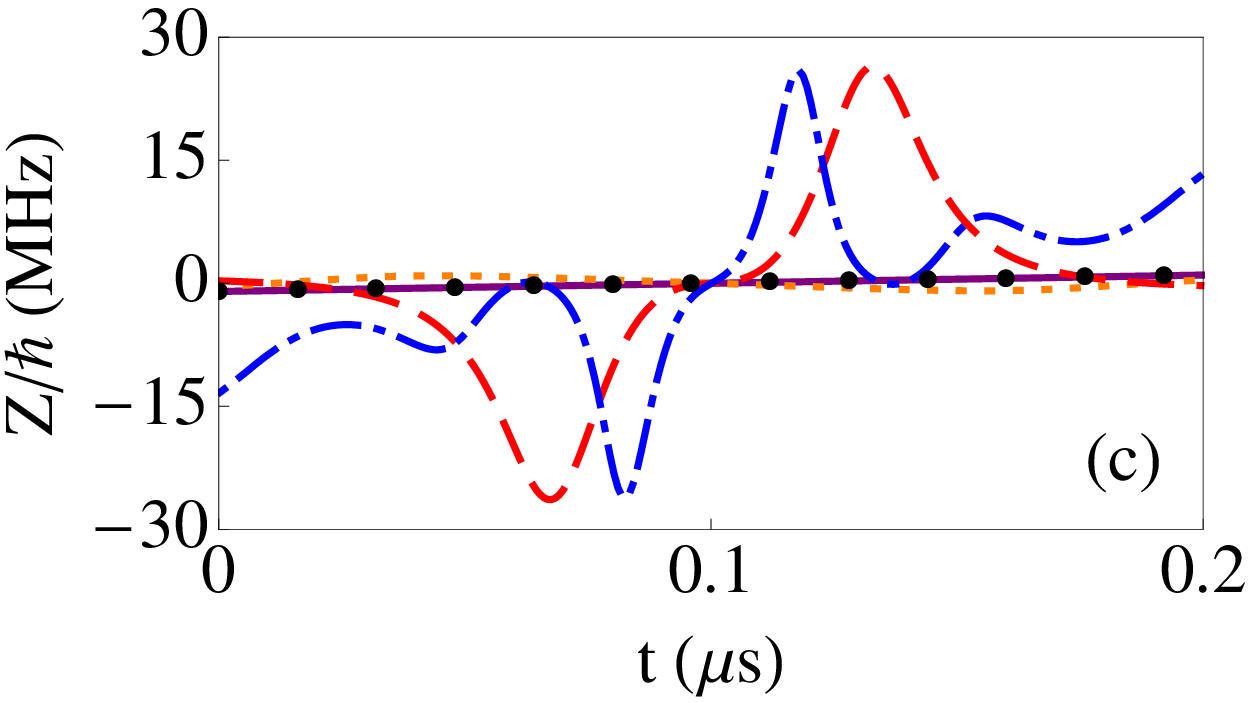}
\caption{\label{H_lz}
(Color online) The (a) $X$, (b) $Y$ and (c) $Z$ components 
for the Landau-Zener scheme, of: $H_0(t)$ (black dots), $H_0^{(1)}$ (purple solid line), $H_0^{(2)}$ (orange dotted line), $H_0^{(3)}$ (red dashed line), and $H_0^{(4)}$ (blue dot-dashed line).
In (a) and (c) the black dots and the green thin line coincide and in (b) the black dots coincide with the orange dotted line.
Parameters: $\alpha=-20$ MHz$^2$, $\Omega_{0,lz}=0.2$ MHz, and $t_f=0.2$ $\mu$s.
}
\end{center}
\end{figure}

As an example we consider now a Landau-Zener scheme for $H_0$ (for the Allen-Eberly scheme we have found similar results), and study the behavior of $H_0^{(j)}$ with $j=1,2,3,4$, and the populations of the bare states driven by these Hamiltonians.
%
%
%
\begin{table}
\caption{Maxima of the X and Y components of $H_0$ and $H_0^{(j)}$ for $j=1,2,3,4,5$.
Parameters: $\alpha=-20$ MHz$^2$,
$\Omega_{0,lz}=0.2$ MHz, and $t_f=0.2$ $\mu$s.}
\begin{center}
\begin{tabular}{@{\hspace{1pt}} c@{\hspace{5pt}} @{\hspace{5pt}} c@{\hspace{5pt}} @{\hspace{5pt}} c @{\hspace{1pt}}}
\hline\hline \\  [-2 ex]
Hamiltonian    &  $|X_{max}|/\hbar$ (MHz)  & $|Y_{max}|/\hbar$ (MHz)   \\  [2ex]
\hline
$H_0/\hbar$  & 0.1  & 0  \\  [1.5 ex]

$H_0^{(1)}/\hbar$  & 0.1  & 49.9  \\  [1.5 ex]

$H_0^{(2)}/\hbar$  & 10  & 0  \\  [1.5 ex]

$H_0^{(3)}/\hbar$  & 8.4  & 2.8  \\  [1.5 ex]

$H_0^{(4)}/\hbar$  & 46.8  & 28.1  \\  [1.5 ex]

$H_0^{(5)}/\hbar$  & 56.2  & 62.8  \\  [2 ex]
\hline
\end{tabular}
\end{center}
\label{table1}
\end{table}
%
%
%
%
%
%
%
%

\begin{figure}[h]
\begin{center}
\includegraphics[height=4.cm,angle=0]{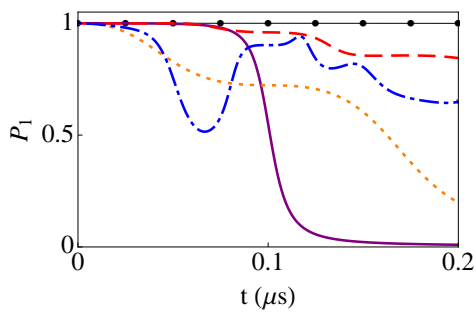}
\caption{\label{P1_lz}
(Color online) Population of the state $|1\rangle$, $P_1(t)$, for the Hamiltonians $H_0(t)$ (black solid line with dots), $H_0^{(1)}$ (purple solid line), $H_0^{(2)}$ (orange dotted line), $H_0^{(3)}$ (red dashed line), and $H_0^{(4)}$ (blue dot-dashed line), with the Landau-Zener scheme. Parameters as in Fig. \ref{H_lz}: $\alpha=-20$ MHz$^2$, $\Omega_{0,lz}=0.2$ MHz, and $t_f=0.2$ $\mu$s.
}
\end{center}
\end{figure}

For the Landau-Zener model $\Delta(t)$ is linear in time and $\Omega_{R}(t)$ is constant, 
\beqa
\label{AE_model}
\Delta_{lz}(t) &=&  \alpha (t-t_f/2),
\nonumber
\\
\Omega_{R, lz}(t) &=& \Omega_{0,lz},
\eeqa
where $\alpha$ is the chirp, and $\Omega_{0,lz}$ is a constant Rabi frequency. 
Condition (\ref{con00}) can be  restated as   
\beq
\label{cond_2}
t_f \gg 2 \left|\frac{\Omega_{0,lz}}{\alpha}\right|.
\eeq
We consider the parameters $\alpha=-20$ MHz$^2$, $\Omega_{0,lz}=0.2$ MHz, 
and $t_f=0.2$ $\mu$s 
for which  the dynamics with $H_0$ is non-adiabatic, see the Appendix A.
Fig. \ref{H_lz} shows $X$, $Y$ and $Z$ components 
of  $H_0$ and $H_0^{(j)}$, with $j=1,2,3,4$. 
In Figs. \ref{H_lz}a and \ref{H_lz}b and in table \ref{table1} we see that 
$H_0^{(2)}$ (corresponding to the first superadiabatic iteration) 
is optimal with respect to applied intensities. 
Moreover it 
cancells the $Y$-component completely, which is    
a simplifying practical advantage in some realizations of the two-level system
\cite{Oliver,Sara12}.  
From the second superadiabatic iteration both intensities start to increase again.
For the parameters above, condition (\ref{cond_2}) is satisfied since $t_f=20 \times |\Omega_{0,lz}/\alpha|$, but not
so condition (\ref{cond_1_0}). 
Fig. \ref{H_lz} shows the disagreement between  $H_0^{(j)}$ 
and $H_0$, at $t=0$ and $t=t_f$ for $j>1$.
Fig. \ref{P1_lz} shows that only $H_0^{(1)}= H_0+ H_{cd}^{(0)}$ inverts the population of $|1\ra$, $P_1(t)$, whereas the rest of the Hamiltonians fail to do so.
 
%
%
%
%
%
%
%
\section{Discussion}
In this paper we have investigated the use of quantum superadiabatic iterations (a non-convergent sequence  of nested interaction pictures) to produce shortcuts to adiabaticity. Each superadiabatic iteration may be used in two ways: ({\it i}) to generate  a superadiabatic approximation to the dynamics, or ({\it ii}) to generate a counterdiabatic term that, when added to the original Hamiltonian, makes the approximate dynamics exact. 
The second approach, however, does not automatically generate shortcuts to adiabaticity, namely, a Hamiltonian that produces in a finite time the same final populations  than the adiabatic dynamics. 
The boundary conditions needed for the second approach to  generate a shortcut have been spelled out. This work is parallel to 
the investigation by Garrido to establish conditions so that  the approach ({\it i})
provides an adiabatic-like approximation \cite{Garrido}. 
We have also described an alternative framework to the usual set of superadiabatic equations
which offers some computational advantages, and have applied the general formalism to the particular case of a two-level system. An optimal superadiabatic iteration with respect to the
norm of the counterdiabatic term, is not necessarily the best shortcut, or in fact a shortcut at all, because of the possible failure of the boundary conditions.    

We end by mentioning further questions worth investigating on the superadiabatic framework 
as a shortcut-to-adiabaticity generator.   
For example, other operations different from 
the population control of two-level systems (such as transport or expansions of cold atoms) 
have to be studied. Unitary transformations may be also applied to simplify the 
Hamiltonian structure making use of symmetries \cite{Sara12}. They  have been  discussed 
before as a way to modify the 
first (adiabatic) iteration \cite{Berry1990,Sara12}, and applied to perform 
a fast population inversion of a
condensate in the bands of an optical lattice \cite{Oliver}, but a systematic
application and study, e.g. of the order with respect to the small (slowness)  parameter, 
in particular for higher superadiabatic iterations, are still pending.      
A comparison with other methods to get shortcuts, at formal and practical levels 
would be useful too. A preliminary step in this direction, relating and comparing the 
invariant-based inverse engineering approach to the counterdiabatic approach 
of the first (adiabatic) iteration was presented in \cite{Inv_Berry}, see also the Appendix B.  
Finally, comparisons among superadiabatic iterations themselves have to be performed, 
in particular regarding practical aspects such as the
transient excitations involved \cite{energy}.

We are grateful to O. Morsch and M. Berry for discussions.   
We acknowledge funding by Projects No. IT472-10, No. FIS2009-12773-C02-01, 
and the UPV/EHU program UFI 11/55.  
S. I. acknowledges  Basque Government Grant No. BFI09.39. X. C. thanks 
the National Natural Science Foundation of China (Grant No. 61176118) and 
Grant No. 12QH1400800.

\appendix
\section{Adiabaticity and boundary conditions for the Landau-Zener protocol} 
\begin{figure}[h]
\begin{center}
\includegraphics[height=4.cm,angle=0]{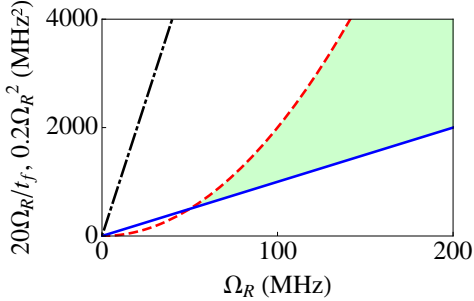}
\caption{\label{conditions}
(Color online) $0.2 \Omega_{0, lz}^2$ (red dashed line) and $20 |\Omega_{0, lz}|/t_f$ (blue dashed line for $t_f=2$ $\mu$s and black dot-dashed line for $t_f=0.2$ $\mu$s). 
The shaded (green) area corresponds to values of $\alpha$ 
satisfying $20 |\Omega_{0, lz}|/t_f  < |\alpha| < 0.2 \Omega_{0, lz}^2$,
namely, the process is adiabatic and the eigenstates at the
boundary times are essentially the bare states.
No such area exists for $t_f=0.2$ $\mu$s
in the domain shown. 
}
\end{center}
\end{figure}

The adiabaticity condition for a two-level atom driven  by 
the Hamiltonian (\ref{H0_2level}) is  \cite{Xi_PRL}
\beq
\frac{1}{2} |\Omega_a(t)| \ll |\Omega(t)|,
\eeq
where $\Omega_a(t) \equiv [\Omega_R(t) \dot{\Delta}(t)-\dot{\Omega}_R(t) \Delta(t)] / \Omega^2(t)$ and $\Omega = \sqrt{\Delta^2 + \Omega_R^2}$. For the  Landau-Zener scheme this condition takes the form
\beq
|\alpha| \ll  2 \Omega_{0, lz}^2.
\label{adiab_cond}
\eeq
%
The inequalities that $\alpha$ must satisfy so that the system is  adiabatic
and also fulfills  the boundary condition (\ref{cond_2}) are  
\beq
2 |\Omega_{0, lz}|/t_f \ll |\alpha| \ll 2 \Omega_{0, lz}^2.
\label{adiab_bc}
\eeq
%
Fig. \ref{conditions} shows the (shaded area) region for which $\alpha$ satisfies 
$20 |\Omega_{0, lz}|/t_f  < |\alpha| < 0.2 \Omega_{0, lz}^2$ 
when $t_f=2$ $\mu$s. No such area exists in the depicted domain for  $t_f=0.2$ $\mu$s. 
For this shorter time the critical point where  
$1/t_f= \Omega_{0, lz}/100$ corresponds to $\Omega_{0, lz} =500$ MHz and 
detunings of up to $5$ GHz. Both may be problematic, as very large laser intensities
and detunings could excite other transitions. 
%
%
%
%
%
\section{Invariants}
The superadiabatic sequence may be pictured as an attempt  
to find a higher order frame for which a coupling term $K_j$
is zero in the dynamical equation so that there are no transitions in some basis.  
This would mean that the states that the system follows exactly have been found,
in other words, the eigenvectors of a dynamical invariant $I(t)$ {\cite{Lewis_Riesenfeld, Lewis_Leach, Inv_Berry}}.
When counter-diabatic terms are added, it is easy to construct invariants
for $H_0^{(j)}$ from the instantaneous eigenstates of $H_0(t)$.
However, quite generally this is not enough to generate a shortcut to adiabaticity because the boundary conditions to perform a quasi-adiabatic process (one that ends up with the same populations than the adiabatic one) may not be satisfied.
A way out is to design the invariant first, and then $H(t)$ from it,
satisfying the boundary conditions
$[I(t),H(t)]=0$ at $t=0$ and $t=t_f$, and such that $H(0)=H_0(0)$ and 
$H(t_f)=H_0(t_f)$ \cite{Inv_Berry,review}. 
\begin{figure}[h]
\begin{center}
\includegraphics[height=4.cm,angle=0]{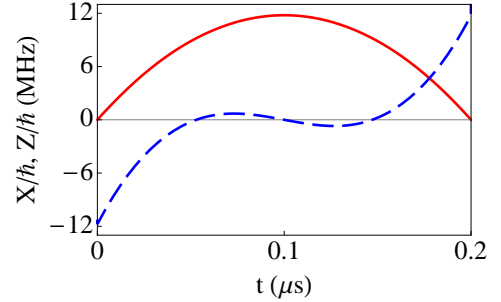}
\caption{\label{Omega_Delta}
(Color online) $X$ (red solid line) and $Z$ (blue dashed line) components of the Hamiltonian 
obtained using the invariant-based inverse engineering method. 
$t_f=0.2$ $\mu$s. 
}
\end{center}
\end{figure}
%
\begin{figure}[h]
\begin{center}
\includegraphics[height=4.cm,angle=0]{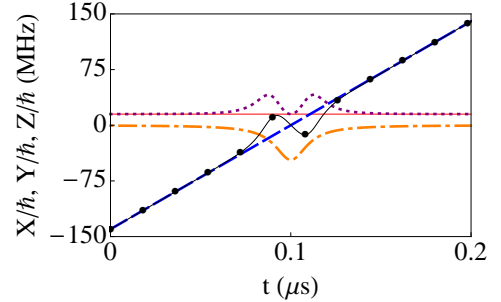}
\caption{\label{Omega1_Delta1}
(Color online) The components $X_{0}(t)$ (red solid line) and $Z_{0}(t)$ (blue dashed line) of $H_0(t)$, the $Y(t)$ component of
$H_0^{(1)}(t)$ (orange dot-dashed line), and the $X(t)$ (purple dotted line) and $Z(t)$ (black solid line with dots) components of $H_0^{(2)}$, for the Landau-Zener scheme. 
Parameters: $\alpha=-2800$ MHz$^2$, $\Omega_{0,lz}=30$ MHz, and $t_f=0.2$ $\mu$s.
}
\end{center}
\end{figure}
%
\begin{figure}[h]
\begin{center}
\includegraphics[height=4.cm,angle=0]{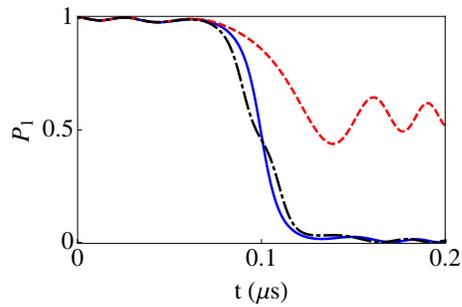}
\caption{\label{p1_p2}
(Color online) Population of $|1\rangle$, $P_1(t)$, for the Hamiltonians $H_0(t)$ (red dashed line), $H_0^{(1)}$ (blue solid line), and
$H_0^{(2)}$ (black dot-dashed line), with the Landau-Zener scheme. Parameters as  in Fig. \ref{Omega1_Delta1}: $\alpha=-2800$ MHz$^2$, $\Omega_{0,lz}=30$ MHz, and $t_f=0.2$ $\mu$s.
}
\end{center}
\end{figure}
For the general Hamiltonian in Eq. (\ref{H0_2level}), 
a dynamical invariant of the corresponding Schr\"odinger
equation  
may be parameterized as \cite{Inv_Berry}
\beqa
\label{I}
I (t)= \frac{\hbar}{2} \nu \left(\begin{array}{cc} \cos{\gamma(t)} & \sin{\gamma(t)} e^{ i \beta(t)}
\\ \sin{\gamma(t)} e^{-i \beta(t)} &  -\cos{\gamma(t)}
\end{array}\right),
\eeqa
where $\nu$ is an arbitrary constant with units of frequency to keep $I(t)$ with dimensions of energy. From the invariance condition for $I$,
\beqa
\frac{dI (t)}{dt} \equiv \frac{\partial{I}(t)}{\partial{t}} - {\frac{i}{\hbar}} [I(t),H_0(t)] = 0,
\eeqa
the functions $\gamma(t)$ and $\beta(t)$
must satisfy the differential equations
\begin{eqnarray}
\dot{\gamma} &=& \Omega_{R}\sin\beta,
\nonumber\\
\dot{\beta} &=& \Delta + \Omega_{R}\cos\beta \cot\gamma.
\label{schrpure}
\end{eqnarray}
To achieve a population inversion, the boundary values for $\gamma$ should be
$\gamma(0)=0$ and $\gamma(t_f)=\pi$. 
Assuming a polynomial ansatz \cite{shortcut_harmonic_traps,transport,Inv_Berry} for $\gamma(t)$ and $\beta(t)$, as $\gamma(t)=\sum_{n=0}^3 a_n t^n$ with the boundary conditions $\gamma(0)=\pi$, $\gamma(t_f)= \dot{\gamma}(0)= \dot{\gamma}(t_f)=0$, and $\beta(t)=\sum_{n=0}^4 b_n t^n$ with the boundary conditions $\beta(0)= \beta(t_f/2)= \beta(t_f)= -\pi/2$,  $\dot{\beta}(t_f)=-\pi/(2t_f)$, and $\dot{\beta}(0)=\pi/(2t_f)$, we can construct $\Delta$ and $\Omega_R$ \cite{Inv_Berry}. These two functions are shown in Fig. \ref{Omega_Delta}, for $t_f=0.2$ $\mu$s ($\Omega_R=2X/\hbar$ and $\Delta=-2Z/\hbar$). 
For the same process time $t_f$ we also plot in {Fig. \ref{Omega1_Delta1} $X_{0}(t)$ and $Z_{0}(t)$
for a Landau-Zener protocol in which the Rabi frequency is slightly larger 
than the maximun required for the invariant-based protocol: $\Omega_{0,lz}=30$ MHz. 
As explained in the previous appendix, an unreasonably high laser intensity would be required to make it adiabatic  while satisfying 
the bare-state condition at the edges, and $\Omega_{0,lz}=30$ MHz is still too small to satisfy 
Eq. (\ref{adiab_bc}). This is evident in the failure to invert the population, 
see Fig. \ref{p1_p2}.   
We use $\alpha=-2800$ MHz$^2$ to have the bare states as eigenvectors at the time edges
which implies a rather large detuning.   
Fig. \ref{Omega1_Delta1} also depicts  
the $Y(t)$ component of $H_0^{(1)}$ and the $X(t)$ and $Z(t)$ components of
$H_0^{(2)}$} for
$t_f=0.2$ $\mu$s. With these parameters these Hamiltonains provide shortcuts to adiabaticity, see Fig. \ref{p1_p2}, but they use very high detunings compared to those of the invariant-based protocol. 
This 
example does not mean, however,  that invariant-based engineering is systematically more efficient. Invariant-based engineering and the counterdiabatic approach provide 
families of protocols  that depend on the chosen interpolating auxiliary functions in the first case and on the reference Hamiltonian $H_0$ in the second. 
Their potential equivalence was studied in  \cite{Inv_Berry}. 
%

%
%

%
\end{document}